\documentclass{jnmp}

\usepackage{amsmath}

\JNMPnumberwithin{equation}{section}
\resetfootnoterule

\newcommand{\wb}{\overline}
\newcommand{\wh}{\widehat}
\newcommand{\wt}{\widetilde}
\newcommand{\cQ}{\mathcal{Q}} 
\newcommand{\cH}{\mathcal{H}}
\newcommand{\ssp}{\mathfrak{p}}
\newcommand{\ssq}{\mathfrak{q}}
\newcommand{\sst}{\mathfrak{t}}
\newcommand{\ssl}{\mathfrak{l}}
\newcommand{\sse}{\mathfrak{e}}
\newcommand{\ssu}{\mathfrak{u}}

\newcommand{\pro}{\cdot}
\newcommand{\map}[1]{\stackrel{#1}{\mapsto}}
\newcommand{\sIm}[1]{\stackrel{#1}{\sim}}
\newcommand{\deform}{_*}

\newtheorem*{acknowledgments}{Acknowledgments}
\theoremstyle{definition}

\begin{document}

\renewcommand{\evenhead}{J Atkinson and F W Nijhoff}
\renewcommand{\oddhead}{Solutions of Adler's lattice equation}

\thispagestyle{empty}

\FirstPageHead{*}{*}{20**}{\pageref{firstpage}--\pageref{lastpage}}{Article}

\copyrightnote{2007}{J Atkinson and F W Nijhoff}

\Name{Solutions of Adler's lattice equation associated with 2-cycles of the B\"acklund transformation}

\label{firstpage}

\Author{James ATKINSON and Frank NIJHOFF}

\Address{Department of Applied Mathematics, University of Leeds, Leeds LS2 9JT, UK, \\
~~E-mail: james@maths.leeds.ac.uk, frank@maths.leeds.ac.uk\\}

\Date{Received Month *, 200*; Accepted in Revised Form Month *, 200*}

\begin{abstract}
\noindent
The B\"acklund transformation (BT) of Adler's lattice equation is inherent in the equation itself by virtue of its multidimensional consistency.
We refer to a solution of the equation that is related to itself by the composition of two BTs (with different B\"acklund parameters) as a {\it 2-cycle} of the BT.
In this article we will show that such solutions are associated with a commuting one-parameter family of rank-2 (i.e., 2-variable), 2-valued mappings.
We will construct the explicit solution of the mappings within this family and hence give the solutions of Adler's equation associated with 2-cycles of the BT.
\end{abstract}


\section{Introduction}

The problem of finding the ``periodic fixed-points'' of a B\"acklund transformation (BT) was first considered by John Weiss \cite{jw1,jw2} in connection with the construction of solutions of the Korteweg-de Vries (KdV) equation.
He obtained finite-dimensional integrable systems associated with such solutions, namely periodic Kac-van Moerbeke chains.

In \cite{ahn} we found solutions of Adler's lattice equation by constructing the \mbox{fixed-points} of its BT.
Adler's equation was discovered in \cite{Adler} as the nonlinear superposition principle for BTs of the Krichever-Novikov (KN) equation \cite{KN1,KN2}, it is an integrable lattice equation in which the lattice parameters are points on an elliptic curve.
This lattice equation is multidimensionally consistent in the sense of \cite{nw,bs}, cf \cite{Nij}, which means the BT is inherent in the lattice equation itself.
Because of this one is tempted to conclude that solutions can therefore be straightforwardly constructed.
However, the construction of a seed solution to start a B\"acklund chain turns out to be a nontrivial problem for Adler's equation, in fact the simplest solution of that equation in terms of elliptic functions is only trivially altered by the BT (we coined such seed solutions \textit{non-germinating}) and leads to a trivial B\"acklund chain.
However the solution found in \cite{ahn} as the fixed-point (or 1-cycle) of the BT yields a nontrivial B\"acklund chain of soliton type solutions of Adler's equation.
Remarkably, this germinating seed solution is again in terms of elliptic functions, but over a deformation of the curve associated with the lattice parameters of the equation.

In the present note we will push this idea one step further and construct solutions of Adler's equation that are 2-cycles of the BT.
It will be seen that such solutions are associated with a commuting one-parameter family of rank-2 (i.e., 2-variable), 2-valued mappings.
The construction we apply is closely related to the issue of periodic reductions of the quadrilateral lattice.
This idea was first explored in the context of periodic ``staircase'' reductions of integrable lattice equations of KdV type, cf \cite{pnc}, where they led to mappings integrable in the sense of Liouville \cite{cnp} (the general notion of Liouville integrability of maps was discussed in \cite{ves}).
A two-step reduction of this type for Adler's equation was studied recently in the work of Joshi et. al. \cite{Joshi} in which the resulting scalar second-order mapping was shown to be of a non-QRT type.
We will explain the connection between this mapping and the mapping associated with the 2-cycle of the BT considered here.
More generally, periodic reductions on the lattice can be considered the analog of finite-gap solutions \cite{en} (cf also \cite{jw1}) and this connection leads us to expect that parameterisation of $N$-cycles of the BT with $N>2$ will need Abelian functions associated with hyper-elliptic curves, which is beyond the scope of this note.

\section{Adler's lattice equation}

We will consider Adler's lattice equation in the Jacobi form, if we define the quadrilateral expression
\begin{equation}
\cQ_{\ssp,\ssq}(u,\wt{u},\wh{u},\wh{\wt{u}}) =  p(u\wt{u}+\wh{u}\wh{\wt{u}})-q(u\wh{u}+\wt{u}\wh{\wt{u}})
-\frac{pQ-qP}{1-p^2q^2}(u\wh{\wt{u}}+\wt{u}\wh{u}-pq(1+u\wt{u}\wh{u}\wh{\wt{u}})),\ \ 
\label{eq:convenient}
\end{equation}
then we may write the equation conveniently as
\begin{equation}
\cQ_{\ssp,\ssq}(u,\wt{u},\wh{u},\wh{\wt{u}}) = 0. 
\label{adler}
\end{equation}
Here $u=u(n,m)$, $\wt{u}=u(n+1,m)$, $\wh{u}=u(n,m+1)$ and $\wh{\wt{u}}=u(n+1,m+1)$ denote the values of the scalar dependent variable $u$ as a function of the independent variables $n,m\in\mathbb{Z}$.
The lattice parameters $\ssp=(p,P)$ and $\ssq=(q,Q)$ lie on the elliptic curve of Jacobi type $\Gamma$,
\begin{equation}
\Gamma = \left\{(x,X):X^2=x^4+1-(k+1/k) \ x^2\right\}
\label{eq:Gamma}
\end{equation}
where $k$ is referred to as the elliptic modulus.
This form of Adler's equation was first given by Hietarinta \cite{Hie}, it is equivalent (by a change of variables) to the Weierstrass form given originally by Adler \cite{Adler}, cf \cite{ahn}.

The natural product that turns $\Gamma$ into an abelian group has the following rational representation
\begin{equation}
\ssp \pro \ssq = \left( \frac{pQ+qP}{1-p^2q^2}, \frac{Pp(q^4-1) - Qq(p^4-1)}{(1-p^2q^2)(qP-pQ)} \right).
\label{gp}
\end{equation}
The identity in this group is the point $\sse = (0,1)$ and the inverse of a point $\ssp$ is the point $\ssp^{-1} = (-p,P)$.
The product (\ref{gp}) will play a central role in our construction of the solutions of (\ref{adler}).
We have defined it here independently of the elliptic modulus $k$, so it can be defined this way for any curve of Jacobi type.
The product (\ref{gp}) is connected to the biquadratic expression
\begin{equation}
\cH_{\ssp}(u,\wt{u}) = \frac{1}{2p}\left( u^2+\wt{u}^2-(1+u^2\wt{u}^2)p^2 - 2u\wt{u}P\right).
\label{hdef}
\end{equation}
Specifically if $U$ is such that $\ssu=(u,U)\in\Gamma$, then $\wt{\ssu} \in \{\ssp\pro\ssu, \ssp^{-1}\pro\ssu\} \ \Rightarrow \ \cH_\ssp(u,\wt{u}) = 0 $ because of the factorisation
\begin{equation}
\cH_{\ssp}(u,\wt{u}) = \frac{1-u^2p^2}{2p}\left(\wt{u}-\frac{uP-pU}{1-p^2u^2}\right)\left(\wt{u}-\frac{uP+pU}{1-p^2u^2}\right).
\label{hfactor}
\end{equation}
In turn the biquadratic expression (\ref{hdef}) is connected to Adler's equation by the identity
\begin{equation}
\cQ_{\ssp,\ssq}(u,\wt{u},\wh{u},\wh{\wt{u}})\cQ_{\ssp,\ssq^{-1}}(u,\wt{u},\wh{u},\wh{\wt{u}}) = \frac{4p^2q^2}{1\!-\!p^2q^2}\left(\cH_\ssp(u,\wt{u})\cH_\ssp(\wh{u},\wh{\wt{u}})-\cH_\ssq(u,\wh{u})\cH_\ssq(\wt{u},\wh{\wt{u}})\right),
\label{fid}
\end{equation}
which arises in Adler's original construction \cite{Adler}.
We shall see that the relation (\ref{fid}) can be used to simplify the construction of solutions presented here.

\section{The B\"acklund transformation and its $2$-cycle}

Let us now describe (and introduce convenient notation for) the BT of Adler's equation.
Consider the coupled system of equations in the variables $u=u(n,m)$ and $v=v(n,m)$,
\begin{equation}
\cQ_{\ssp,\ssl}(u,\wt{u},v,\wt{v}) = 0, \qquad \cQ_{\ssq,\ssl}(u,\wh{u},v,\wh{v})=0.\\
\label{bt}
\end{equation}
These equations involve shifts on $u$ and $v$ in the single discrete variable $n$ and $m$ respectively, so they are ordinary difference equations.
If the pair $u$,$v$ satisfy (\ref{bt}) throughout the lattice (i.e., for all $n$ and $m$) then we will write 
\begin{equation} u \sIm{\ssl} v \label{btrel} \end{equation}
and say that $v$ is related to $u$ by the B\"acklund transformation (\ref{bt}) with B\"acklund parameter $\ssl$.
By the symmetry of the expression (\ref{eq:convenient}), the system (\ref{bt}) is invariant under the interchange $u \leftrightarrow v$, so the relation (\ref{btrel}) is symmetric.
Now, if $u$ is fixed throughout the lattice, then (\ref{bt}) constitutes an overdetermined system for $v$, it can be verified that that this system is compatible if and only if $u$ satisfies (\ref{adler}).
Together with the $u\leftrightarrow v$ symmetry this implies that if $u\sIm{\ssl}v$, then both $u$ and $v$ satisfy (\ref{adler}).

In the communication \cite{ahn} we considered the solutions of (\ref{adler}) that are \mbox{fixed-points} of the BT (\ref{bt}), more precisely these are solutions for which \begin{equation}u \sIm{\sst} u\end{equation} for some fixed parameter $\sst\in\Gamma$.
This is the simplest case of the more general problem to find the ``periodic \mbox{fixed-points}'' of the BT in the sense of Weiss \cite{jw1,jw2}.
These are solutions $u_1 \ldots u_N$ for which
\begin{equation}
u_1 \sIm{\sst_1} u_2, \qquad u_2 \sIm{\sst_2} u_3, \qquad \ldots \qquad u_N \sIm{\sst_N} u_1
\label{ncycle}
\end{equation}
for some parameters $\sst_1 \ldots \sst_N \in \Gamma$.
That such solutions of (\ref{adler}) exist is not a-priori obvious, however it is worth remarking that the system of equations implied by (\ref{ncycle}) amounts to a coupled pair of rank-$N$ 2-valued mappings, the commutativity of which is equivalent to the existence of these solutions.
  
In the present article we will consider this problem in the case $N=2$, that is to find $u$ for which there exists $v$ such that
\begin{equation} u \sIm{\sst_1} v, \qquad v \sIm{\sst_2} u, \label{1bituv} \end{equation}
for some fixed parameters $\sst_1,\sst_2 \in \Gamma$.
If $\sst_1=\sst_2$ this includes {\it all} solutions of (\ref{adler}) because the BT relation is symmetric.
If $\sst_1\neq \sst_2$ we will refer to any such $u$ as a {\it 2-cycle} of the BT.
Clearly the solution $v$ is also a 2-cycle of the BT.
By definition, the equations implied by (\ref{1bituv}) are
\begin{eqnarray}
\cQ_{\ssp,\sst_1}(u,\wt{u},v,\wt{v}) = 0, &\qquad& \cQ_{\ssp,\sst_2}(u,\wt{u},v,\wt{v})=0,\label{sys}\\
\cQ_{\ssq,\sst_1}(u,\wh{u},v,\wh{v}) = 0, &\qquad& \cQ_{\ssq,\sst_2}(u,\wh{u},v,\wh{v})=0.\label{1o2}
\end{eqnarray}
We will treat the systems (\ref{sys}) and (\ref{1o2}) as mappings $(u,v) \map{\ssp} (\wt{u},\wt{v})$ and $(u,v) \map{\ssq} (\wh{u},\wh{v})$ respectively.
As such they are 2-valued, that is if we fix $(u,v)$ then there are two possible values of the pair $(\wt{u},\wt{v})$ which satisfy (\ref{sys}), and two possible values of the pair $(\wh{u},\wh{v})$ which satisfy (\ref{1o2}).
In order to construct the general simultaneous solution of these mappings we begin in the following section by solving the first of them, the system (\ref{sys}).

\section{The explicit solution of the rank-2, 2-valued mapping and the deformed elliptic curve}

Consider the mapping $(u,v) \map{\ssp} (\wt{u},\wt{v})$ defined by the system (\ref{sys}).
This involves shifts in the discrete variable $n$ only, hence throughout this section, in which we solve (\ref{sys}), we will restrict our attention to the variable $n$ alone, that is we consider \mbox{$(u,v)=(u(n),v(n))$}.
In the first step toward the solution of this mapping we use the identity (\ref{fid}), which shows that the defining equations (\ref{sys}) imply that
\begin{equation}
\begin{array}{rcl}
\cH_\ssp(u,\wt{u})\cH_\ssp(v,\wt{v}) &=& \cH_{\sst_1}(u,v)\cH_{\sst_1}(\wt{u},\wt{v}), \\
\cH_\ssp(u,\wt{u})\cH_\ssp(v,\wt{v}) &=& \cH_{\sst_2}(u,v)\cH_{\sst_2}(\wt{u},\wt{v}).
\end{array}
\label{dsys}
\end{equation}
Elimination of the common LHS from the derived system (\ref{dsys}) leads naturally to the following
\begin{equation} J\wt{J}=1, \qquad J = \frac{\cH_{\sst_1}(u,v)}{\cH_{\sst_2}(u,v)}. \label{Jdef} \end{equation}
The dynamical equation for the new variable $J$ is trivial and provides a first integral of the derived mapping defined by (\ref{dsys}).
In the remainder of this section we solve the essentially technical problem of using this to find the explicit solution of (\ref{sys}).

Fixing $J(0)$ from the initial data $(u(0),v(0))$, the second equation of (\ref{Jdef}) amounts to a mildly non-autonomous biquadratic constraint on $u$ and $v$ which is actually of Jacobi type:
\begin{equation} \cH_{\sst_2}(u,v)J - \cH_{\sst_1}(u,v) \ \ = \ \ t\deform\frac{t_2-t_1J}{t_1t_2}\ \cH_{\sst\deform}(u,v), \label{defcon} \end{equation}
the new parameter $\sst\deform=(t\deform,T\deform)$ is defined by the equations
\begin{equation} t\deform^2 = t_1t_2\frac{t_1-t_2J}{t_2-t_1J}, \quad T\deform = \frac{t_2T_1-t_1T_2J}{t_2-t_1J} \label{defpar} \end{equation}
and lies on a new curve, $\sst\deform \in \Gamma\deform$,
\begin{equation} \Gamma\deform = \{ (x,X): X^2=x^4+1-(k\deform+1/{k\deform}) x^2 \}, \label{defG} \end{equation}
\begin{equation} k\deform + \frac{1}{k\deform} = \frac{t_1t_2\left(k+1/k\right)\left(J+1/J\right) + 2\left(T_1T_2-1-t_1^2t_2^2\right)}{t_1t_2(J+1/J)-t_1^2-t_2^2}, \label{defk} \end{equation}
which is again of Jacobi type but with a new elliptic modulus $k\deform$.
The parameter $\sst\deform \in \Gamma\deform$ defined by (\ref{defpar}) depends on $J$ so it is non-autonomous, the curve itself depends on $J$ only through the combination $J+1/J$ which from (\ref{Jdef}) is clearly autonomous, it follows that $\wt{\Gamma}\deform=\Gamma\deform$.

The biquadratic constraint $\cH_{\sst\deform}(u,v)=0$ can be used to eliminate $v$ from (\ref{sys}).
We begin by choosing $U$ such that $\ssu = (u,U) \in \Gamma\deform$ and assuming one factor (see identity (\ref{hfactor})) of $\cH_{\sst\deform}(u,v)$ is zero:
\begin{equation} v = \frac{uT\deform + t\deform U}{1-t\deform^2u^2}. \label{uv} \end{equation}
This is without loss of generality because choosing the other factor leads to (\ref{uv}) with $t\deform \rightarrow -t\deform$, under which the equations (\ref{defpar}) defining $\sst\deform$ are invariant.
Note that the relation (\ref{uv}) applied at $n=0$, taken with the definition of $\sst\deform$ (\ref{defpar}), fixes $t\deform$ uniquely at the origin in terms of $\ssu(0) \in \Gamma$ and $v(0)$ which we take as the initial conditions.

On substituting for $v$ using (\ref{uv}), the system (\ref{sys}) reduces to two relations between $\ssu$ and $\wt{\ssu}$.
It turns out that these relations are compatible only if
\begin{equation} t\deform\wt{t}\deform + t_1 t_2 = 0. \label{constraint} \end{equation}
It can be confirmed that this constraint is compatible with the definition of $\sst\deform$, in fact (\ref{constraint}) refines this definition by fixing $\sst\deform$ uniquely at each iteration in terms of its previous value.
Note also that $\wt{\wt{\sst}}\deform=\sst\deform$ so the value of the parameter oscillates.

When (\ref{constraint}) holds the substitution of (\ref{uv}) reduces (\ref{sys}) to a single equation on the curve $\Gamma\deform$,
\begin{equation} \wt{\ssu} = \ssp\deform \pro \ssu, \label{sm} \end{equation}
where the new (non-autonomous) parameter $\ssp\deform=(p\deform,P\deform) \in \Gamma\deform$ is defined by the equations
\begin{equation}
\begin{array}{l}
\begin{displaystyle}\frac{p\deform^2-pp_{12}}{p\deform(p-p_{12})} = \frac{t\deform^2+t_1t_2}{t\deform(t_1+t_2)}\end{displaystyle},\\
\begin{displaystyle}P\deform = \frac{1}{t_{12}}\left(p-p_{12}\right) + \frac{p\deform^2-pp_{12}}{p-p_{12}}\left(\frac{T_1-T_2}{t_1-t_2}-\frac{t_1-t_2}{p_1-p_2}(p-p_{12})p_1p_2\right)\end{displaystyle}.
\end{array}
\label{pdef}
\end{equation}
We have used the notation: 
\[ \ssp_1 = \ssp \pro \sst_1^{-1}, \quad \ssp_2 = \ssp \pro \sst_2^{-1}, \quad \sst_{12} = \sst_1 \pro \sst_2, \quad \ssp_{12}=\ssp\pro\sst_{12}^{-1} \]
where $\ssp_1=(p_1,P_1)$ etc.
The equations (\ref{pdef}) for $\ssp\deform$ have two solutions, so there is some choice in the parameter $\ssp\deform$ at each iteration of (\ref{sm}).
In fact
\begin{equation} \wt{\ssp}\deform \in \{ \ssp\deform^{-1}, \sst^{-1}\deform \pro \wt{\sst}\deform \pro \ssp\deform \}, \label{choice} \end{equation}
which can be verified directly.
The existence of this choice in the value of $\ssp\deform$ is a consequence of the underlying mapping defined by (\ref{sys}) being 2-valued.
Apart from the book-keeping involved in this detail, the solution of the dynamical equation defined by (\ref{sm}) and (\ref{choice}) is trivial.

We conclude this section by giving a concrete example.
Let us define the {\it canonical} solution by choosing from (\ref{choice}) $\wt{\ssp}\deform = \sst^{-1}\deform \pro \wt{\sst}\deform\pro\ssp\deform$, which fixes $\ssp\deform$ at each iteration in terms of its value at the origin, $\ssp\deform(0)$, moreover $\wt{\wt{\ssp}}\deform=\ssp\deform$ so the value of $\ssp\deform$ oscillates.
Now from (\ref{sm})
\begin{equation}
\ssu(n) = \left\{ \begin{array}{ll}
\ssp\deform(0)^{n/2}\pro\ssp\deform(1)^{n/2} \pro \ssu(0),& n \textrm{ even},\\
\ssp\deform(0)^{(n+1)/2}\pro\ssp\deform(1)^{(n-1)/2}\pro\ssu(0),& n \textrm{ odd}, 
\end{array} \right.
\label{consol}
\end{equation}
where $\ssp\deform(1) = \sst\deform(0)^{-1}\pro\sst\deform(1)\pro\ssp\deform(0)$.
The canonical solution of (\ref{sys}) itself is actually the pair $(u(n),v(n))$, but $v(n)$ can also be found from $\ssu(n)=(u(n),U(n))$ given in (\ref{consol}) by using the relation (\ref{uv}).

\section{The solution on the lattice}

As we have established, any 2-cycle of the BT satisfies the coupled systems (\ref{sys}) and (\ref{1o2}).
In the previous section we solved the first system (\ref{sys}) by reducing it to a single equation on the new curve $\Gamma\deform$.
The second system, defined by (\ref{1o2}), differs from the first only in the change of parameter $\ssp\rightarrow \ssq$ (and that it involves shifts in the other direction, the discrete variable $m$ and not $n$).
In particular the (mildly non-autonomous) biquadratic constraint is the same: $\cH_{\sst\deform}(u,v)=0$, which therefore holds throughout the lattice.
We can use the same substitution (\ref{uv}) to eliminate $v$ from (\ref{1o2}) which, provided $\wh{t}\deform t\deform + t_1t_2=0$ (note from (\ref{constraint}) we see that $\wh{\sst}\deform = \wt{\sst}\deform$ so that $\sst\deform$ is a function of $n+m$ only) then reduces to the single equation
\begin{equation}  \wh{\ssu} = \ssq\deform \pro \ssu, \label{sm2} \end{equation}
on the curve $\Gamma\deform$.
The new parameter $\ssq\deform$ is defined by the relations (\ref{pdef}) with the change $\ssp\rightarrow\ssq$ and $\ssp\deform \rightarrow \ssq\deform$, and it satisfies the dynamical equation
\begin{equation} \wh{\ssq}\deform \in \{ \ssq^{-1}\deform, \sst^{-1}\deform \pro \wh{\sst}\deform \pro \ssq\deform \}. \label{qdef} \end{equation}

It remains to couple the mappings (\ref{sm}) and (\ref{sm2}) to find the full solution on the lattice.
These mappings are compatible, i.e., $\wt{\wh{\ssu}} = \wh{\wt{\ssu}}$, if and only if
\begin{equation} \wh{\ssp}\deform \pro \ssq\deform = \wt{\ssq}\deform \pro \ssp \deform \label{compat} \end{equation}
throughout the lattice.
Now, from the definition of $\ssp\deform$ together with the observation that $\wt{\sst}\deform = \wh{\sst}\deform$, we see that $\wh{\ssp}\deform$ is subject (in principle) to the same choice as $\wt{\ssp}\deform$ in (\ref{choice}).
Similarly $\wt{\ssq}\deform$ is subject to the same choice as $\wh{\ssq}\deform$ in (\ref{qdef}).
However the condition (\ref{compat}) constrains these choices, specifically we must choose that
\begin{equation} 
\wh{\ssp}\deform = \sst^{-1}\deform \pro \wh{\sst}\deform \pro \ssp\deform, \qquad
\wt{\ssq}\deform = \sst^{-1}\deform \pro \wt{\sst}\deform \pro \ssq\deform. \label{cp} \end{equation}
So in fact the dynamics of $\ssp\deform$ in the $\wh{\phantom{u}}$ direction and $\ssq\deform$ in the $\wt{\phantom{u}}$ direction are single-valued.
When the dynamics of $\ssp\deform$ and $\ssq\deform$ satisfy (\ref{cp}) the equations (\ref{sm}) and (\ref{sm2}) can be coupled and the full solution on the lattice, $u(n,m)$, follows from their general simultaneous solution $\ssu(n,m)$.

It is natural to define the {\it canonical 2-cycle of the BT} by fixing the choices (\ref{choice}) and (\ref{qdef}) so that
\begin{equation}
\wt{\ssp}\deform = \ssp\deform \pro \wt{\sst}\deform \pro \sst^{-1}\deform, \quad \quad \wh{\ssq}\deform = \ssq\deform \pro \wh{\sst}\deform\pro \sst^{-1}\deform.
\end{equation}
Given (\ref{cp}) this means that $\wh{\ssp}\deform=\wt{\ssp}\deform$ and $\wh{\ssq}\deform=\wt{\ssq}\deform$ throughout the lattice, so both parameters are a function of $n+m$ only, moreover they oscillate, $\wt{\wt{\ssp}}\deform=\ssp\deform$ etc.
This canonical solution written explicitly is
\begin{equation}
\ssu(n,m) = \left\{ \begin{array}{ll}
\sst\deform(0)^{-(n+m)/2}\pro\sst\deform(1)^{(n+m)/2}\pro\ssp\deform(0)^{n}\pro\ssq\deform(0)^{m}\pro\ssu(0),& n+m \textrm{ even},\\
\sst\deform(0)^{-(n+m-1)/2}\pro\sst\deform(1)^{(n+m-1)/2}\pro\ssp\deform(0)^{n}\pro\ssq\deform(0)^{m}\pro\ssu(0),& n+m \textrm{ odd}. 
\end{array} \right.
\label{lconsol}
\end{equation}
We finish this section with a number of remarks regarding the obtained results.
\subsection*{Remark 1}
Choosing initial data so that $J$ defined in (\ref{Jdef}) is equal to $1$ at the origin has the consequence that $J=1$ throughout the lattice.
Fixing $J=1$ in (\ref{Jdef}) and using this to eliminate $v$ from the systems (\ref{sys}) and (\ref{1o2}) which define the 2-cycle of the BT, yields the reduced system
\begin{equation}
\begin{array}{l}
\cQ_{\ssp,\sst_1\pro\sst_2}(u,\wt{u},u,\wt{u}) = 0,\\
\cQ_{\ssq,\sst_1\pro\sst_2}(u,\wh{u},u,\wh{u}) = 0.
\end{array}
\end{equation}
But this is exactly the system that defines the fixed-point of the BT:
\begin{equation} u \sIm{\sst_1\pro\sst_2} u, \end{equation}
where the B\"acklund parameter associated to this solution is the point $\sst_1\pro\sst_2\in\Gamma$.
So the solution found in this article as a 2-cycle of the BT is a generalisation of the solution found in \cite{ahn} as a fixed-point of the BT because it reduces to that solution if we choose the initial data so that $J=1$.
\subsection*{Remark 2}
In the limit $\sst_2 \longrightarrow \sst_1^{-1}$ we find that $\Gamma\deform \longrightarrow \Gamma$ and $\ssp\deform,\ssq\deform \longrightarrow \ssp,\ssq$.
In this sense the new curve and parameters are deformations of the original curve and lattice parameters associated to the equation (\ref{adler}).
In the same limit the solution presented here goes to the non-germinating seed solution given in \cite{ahn}.
\subsection*{Remark 3}
The superposition formula for solutions of Adler's equation (\ref{adler}) that are related by the BT (\ref{bt}) is inherent in the equation itself:
if $u \sIm{\ssl_1} v$ and $u \sIm{\ssl_2} w$, then $\wb{u}$ defined by the equation $\cQ_{\ssl_1,\ssl_2}(u,v,w,\wb{u}) = 0$ satisfies $v \sIm{\ssl_2} \wb{u}$ and $w \sIm{\ssl_1} \wb{u}$.

Now, given that the 2-cycle of the BT is defined by the relations $u\sIm{\sst_1}v$, $u\sIm{\sst_2}v$, we can naturally construct a new solution $\wb{u}$ by superposition,
\begin{equation} \cQ_{\sst_1,\sst_2}(u,v,v,\wb{u}) = 0, \label{sup} \end{equation}
so that $v\sIm{\sst_2}\wb{u}$ and $v\sIm{\sst_1}\wb{u}$, and hence $\wb{u}$ is another 2-cycle of the BT.
Clearly by iteration of this procedure we can construct a sequence of such solutions.
Note however that solutions related by superposition are associated to the same deformed curve $\Gamma\deform$.

\subsection*{Remark 4}
Let us restrict our attention to the mapping defined by (\ref{sys}) in the special case that $\ssp=\sst_1$, that is we choose one lattice parameter of the equation to coincide with one of the B\"acklund parameters.
In this case the first equation of (\ref{sys}) reduces to the trivial equation $(\wt{u}-v)(\wt{v}-u) = 0$.
Choosing the solution $v=\wt{u}$ brings the second equation of (\ref{sys}) to
\begin{equation} \cQ_{\sst_1,\sst_2}(u,\wt{u},\wt{u},\wt{\wt{u}})=0. \label{red} \end{equation}
This scalar second-order ordinary difference equation is the two-step periodic ``staircase'' reduction of Adler's equation considered first by Joshi et. al. \cite{Joshi}.
Note that, apart from notational differences, the equation (\ref{red}) coincides with the superposition formula for 2-cycles of the BT, (\ref{sup}).
The first integral of the mapping defined by (\ref{sys}) given in (\ref{Jdef}) is unchanged in the case $\ssp=\sst_1$ because it is independent of the parameter $\ssp$.
This integral was first given for (\ref{red}) in the work of Joshi et. al. \cite{Joshi}.

Now, it is of some interest to consider the solution of (\ref{red}) in its own right.
The solution method of section 4 applied in this case leads to (\ref{sm}) being simplified to the (single-valued) mapping $\wt{\ssu} = \sst\deform \pro\ssu$.
The solution itself can be written explicitly as
\begin{equation}
\ssu(n) = \left\{ \begin{array}{ll}
\sst\deform(0)^{n/2}\pro\sst\deform(1)^{n/2} \pro \ssu(0),& n \textrm{ even},\\
\sst\deform(0)^{(n+1)/2}\pro\sst\deform(1)^{(n-1)/2}\pro\ssu(0),& n \textrm{ odd}.
\end{array} \right.
\label{dconsol}
\end{equation}
Note that the parameter $\sst\deform$ that was defined originally in terms of $(u(0),v(0))$ is now defined in terms of $(u(0),u(1))$ because we have chosen $v=\wt{u}$, so that in particular, $v(0)=u(1)$.

\subsection*{Remark 5}

Adler's equation \cite{Adler} is the superposition principle for BTs of the Krichever-Novikov (KN) equation \cite{KN1,KN2}.
In Jacobi form the KN equation is 
\begin{equation} u_y - u_{xxx} + \frac{3}{2u_x}\left(u_{xx}^2-u^4-1+(k+1/k)u^2\right) = 0, \label{kn} \end{equation}
where $x,y \in \mathbb{R}$ are the independent variables.
The BT for (\ref{kn}) found in \cite{Adler} can be written 
\begin{equation} u_x v_x = \cH_\ssl(u,v) \label{knbt} \end{equation}
where $\ssl$ is the B\"acklund parameter.

Now, the equations that define the 2-cycle of the BT (\ref{knbt}) are
\begin{equation}
u_xv_x = \cH_{\sst_1}(u,v), \qquad u_xv_x = \cH_{\sst_2}(u,v). \label{csys}
\end{equation}
This actually constrains the initial data, i.e., by eliminating derivatives we see that $\cH_{\sst_1}(u,v)=\cH_{\sst_2}(u,v)$ which fixes $v$ in terms of $u$ for all $x$.
In fact eliminating $v$ from (\ref{csys}) the resulting equation for $u$ is
\begin{equation} u_x^2 = \cH_{\sst_1 \pro \sst_2}(u,u).\label{cont} \end{equation}
This is nothing but the defining equation for the fixed-point of the BT of the KN equation (the seed solution given in \cite{ahn}) with B\"acklund parameter $\sst_1\pro\sst_2 \in \Gamma$.
Therefore for the KN equation the 2-cycle of the BT and the fixed-point of the BT essentially coincide.

\section{Conclusion}

The solutions of Adler's lattice equation found in this article as 2-cycles of the B\"acklund transformation (BT) have been shown to generalise the solutions we found previously as \mbox{fixed-points} (or 1-cycles) of the BT.
Like those solutions, the 2-cycles of the BT are in terms of shifts on a deformation of the elliptic curve associated with the equation itself.
The new features in this case are that the deformation of the curve depends on the choice of initial data, and the shifts on the deformed curve are themselves non-autonomous (in fact they oscillate with period 2).

In the continuous case, i.e., for the Krichever-Novikov equation, we have found that the solutions which arise as 2-cycles of the BT essentially coincide with the solutions we found previously as \mbox{fixed-points} of the BT.
This is in contrast to the situation for Adler's equation and reveals a notable difference between the compatible discrete and continuous systems.

In this article we have also outlined the more general construction of $N$-cycles of the BT for Adler's equation, of which the 1-cycles and 2-cycles discussed above are special cases.
It is natural to conjecture that the rank-$N$ 2-valued mapping that arises in this more general case is integrable.

\begin{acknowledgments}
JA was supported by the UK Engineering and Physical Sciences Research Council (EPSRC).
Both authors would like to thank the organisers of NEEDS 2007.
We would also like to thank Jarmo Hietarinta for useful discussions.
\end{acknowledgments}

\label{lastpage}

\end{document}